\documentstyle[sprocl]{article}

\newcommand{\beq}{\begin{equation}}
\newcommand{\eeq}{\end{equation}}
\newcommand{\beaq}{\begin{eqnarray}}
\newcommand{\eeaq}{\end{eqnarray}}

\newcommand{\xbj}{x_{\scriptscriptstyle B}}
\newcommand{\bpt}{\bm p_\sT}

\newcommand{\psibar}{\overline{\psi}}

\newcommand{\slsh}[1]{\mbox{$\not\! #1$}}
\newcommand{\bm}[1]{\mbox{\boldmath${#1}$}}

\newcommand{\sT}{{\scriptscriptstyle T}}
\newcommand{\sL}{{\scriptscriptstyle L}}

\begin{document}

\title{
THEORETICAL ASPECTS OF AZIMUTHAL AND TRANSVERSE SPIN
ASYMMETRIES\footnote{
Talk at the 9th International Workshop on Deep Inelastic Scattering and QCD
(DIS2001), Bologna, Italy, Apr. 27- May 1, 2001}
}

\author{
\underline{P.J.\ MULDERS}$^1$, A.A. HENNEMAN$^1$, and D. BOER$^2$}
 
\address{\mbox{}\\
$^1$Division of Physics and Astronomy, Vrije Universiteit \\
De Boelelaan 1081, NL-1081 HV Amsterdam, the Netherlands
\mbox{}\\
$^2$RIKEN-BNL Research Center\\
Brookhaven National Laboratory, Upton, NY 11973, U.S.A.\\
}


\maketitle\abstracts{
We use Lorentz invariance and the QCD equations of motion to study
the evolution of functions that appear at leading (zeroth) order in a $1/Q$
expansion in azimuthal asymmetries. 
This includes the evolution equation of the Collins fragmentation function. 
The moments of these functions are matrix elements of known twist two 
and twist three operators.
We present the evolution in the large $N_c$ limit, restricted to
the non-singlet case for the chiral-even functions.
}

In this contribution I want to present one possible way to investigate the
QCD evolution of azimuthal asymmetries~\cite{HBM}.
These asymmetries appear in hard scattering processes with at least two
relevant hadrons and constitute a rich phenomenology, suitable for studying
quark and gluon correlations in hadrons. By relevant hadrons we mean hadrons
used as target or detected in the final state. A well-known azimuthal asymmetry
appears in  the semi-inclusive deep inelastic polarized leptoproduction of
pions ($e p^\uparrow \to e' \pi X$) generated by the so-called Collins 
effect~\cite{Collins-93b}. This asymmetry is one of
the possibilities to gain access to the so-called transversity or
transverse spin distribution function \cite{RS-79,Jaffe-Ji-92}, which is
the third distribution function needed for the complete characterization of the
(collinear) spin state of a proton as probed in hard scattering processes. 
In contrast to the transversity function, the evolution of the Collins 
fragmentation function had not been investigated sofar. Knowledge of this
evolution is indispensable for relating measurements at different energies. 
              
For azimuthal asymmetries~\cite{examples} in processes like semi-inclusive
leptoproduction, often appearing coupled to the spin of the partons and/or
hadrons, it is  important to take transverse momentum of partons into
account, first studied by Ralston and Soper~\cite{RS-79} for the Drell-Yan
process at tree level. Also the Collins function involves transverse
momenta. Furthermore, it is a socalled T-odd function
allowed because time-reversal symmetry does not pose constraints for
fragmentation functions. Its evolution will be one of the new results
presented here, although we limit ourselves to the large $N_c$ limit, in which
case the evolution for T-odd $p_\sT$-dependent functions is autonomous.
  
Factorization crucially depends on the presence of a large energy scale in 
the process, such as the space-like momentum transfer squared $q^2 = -Q^2$ in 
leptoproduction. In this paper we will be concerned with functions that
appear in processes which have, apart from such a hard scale, an
additional soft momentum scale, related to the transverse momentum 
of the partons. In one-hadron inclusive leptoproduction this scale appears
because one deals with three momenta: the large momentum transfer $q$, the
target momentum $P$ and the momentum of the produced hadron $P_h$. The
noncollinearity at the quark level appears via $q_\sT$ = $q + \xbj\,P -
P_h/z_h$, where $\xbj =  Q^2/2P\cdot q$ and $z_h = P\cdot P_h/P\cdot q$ are
the usual semiinclusive scaling variables, at large $Q^2$ identified with
lightcone momentum fractions. The hadron momenta $P$ and $P_h$ define in
essence two lightlike directions $n_+$ and $n_-$, respectively. The soft scale
is $Q_\sT^2 = -q_\sT^2$.

To study the scale dependence of the various distribution and fragmentation
functions appearing in these (polarized) processes  we construct specific
moments in both $p_\sT$ and $x$, employ Lorentz  invariance and use the QCD
equations of motion. The moments in $x$ for leading (collinear)  distribution
functions (appearing for instance in inclusive leptoproduction) are related to
matrix  elements of twist two operators. On the other hand, for the transverse 
moments entering the azimuthal asymmetry expressions of interest, one finds 
relations to matrix elements of twist two {\em and\/} twist three operators, 
for which the evolution, however, is known. In the large $N_c$ limit this 
evolution becomes particularly simple.

In hard processes the effects of hadrons can be studied via quark and gluon
correlators. In inclusive deep inelastic scattering (DIS), these are lightcone
correlators depending on $x \equiv p^+/P^+$ of the type
\beaq
\Phi_{ij} (x) &\equiv& \left. \int \frac{d \xi^-}{2\pi}\ e^{i\,p\cdot
\xi}\langle P,S|\, \psibar_j (0) \,{\cal U}(0,\xi)
\,\psi_i(\xi) | P,S\rangle\right|_{LC} .
\label{PhiDIS}
\eeaq
where the subscript `LC' indicates $\xi^+ = \xi_\sT = 0$ and
${\cal U}(0,\xi)$ is a gauge link with the path running along the minus
direction. The parametrization relevant for DIS at leading (zeroth) order in a
$1/Q$ expansion is 
\beq
\Phi^{{\rm twist}-2}(x)=\frac{1}{2} \left\{{f_1(x)} \slsh n_+ 
+ S_\sL\,{g_1(x)}\,\gamma_{5}\slsh n_+
+ {h_1(x)}\,\gamma_{5}\slsh S_\sT\slsh n_+ \right\} ,
\eeq
where longitudinal spin $S_\sL$ refers to the component along the same
lightlike direction as defined by the hadron. Specifying also the flavor one
also encounters the notations $q(x) = f_1^q(x)$,  $\Delta q(x) = g_1^q(x)$ and
$\delta q(x) = \Delta_T q(x) = h_1^q(x)$. The evolution equations for these
functions are known to next-to-leading order and for the singlet $f_1$ and
$g_1$ there is mixing with the unpolarized and polarized gluon distribution
functions $g(x)$ and $\Delta g(x)$, respectively. 

For DIS up to order $1/Q$ one needs also the 
$M/P^+$ parts in the parameterization of $\Phi(x)$,
\beaq
\Phi^{{\rm twist}-3}(x) &=& \frac{M}{2P^+} \left\{{e(x)} 
+ {g_T(x)}\,\gamma_{5}\slsh S_\sT
+ S_\sL\,{h_L(x)}\,\gamma_{5}\frac{[\slsh n_+,\slsh n_-]}{2} \right\} 
\nonumber \\ 
& + & \frac{M}{2P^+} \left\{-i\,S_\sL\,{e_L(x)} \gamma_5
- {f_T(x)}\,\epsilon_\sT^{\rho\sigma}\gamma_\rho S_{\sT\sigma}
+ i\,{h(x)}\frac{[\slsh n_+,\slsh n_-]}{2} \right\}. 
\eeaq
We have not imposed time-reversal invariance in order to study also the
T-odd functions, which are particularly important in the study of
fragmentation.  The functions $e$, $g_T$ and $h_L$ are T-even, the
functions $e_L$, $f_T$ and $h$ are T-odd. 
The leading order evolution of $e$, $g_T$ and $h_L$
is known \cite{Evol} and for the non-singlet
case this also provides the evolution of the T-odd functions $e_L$,
$f_T$ and $h$ respectively, for which the operators involved differ only from
those of the T-even functions by a $\gamma_5$ matrix. 
The twist assignment is more evident by connecting these functions to the
Fourier transforms of matrix elements of the form $\langle P,S \vert \overline
\psi_j(0)\,{\cal U}(0,\eta) \,iD_\sT^\alpha (\eta)\,{\cal
U}(\eta,\xi)\,\psi_i(\xi) \vert P,S \rangle$ via
the QCD equations of motion.

For a semi-inclusive hard scattering process in which two 
hadrons are identified (in either initial or final state)
the treatment of transverse momentum is important. Instead of lightcone
correlations one needs lightfront correlations (where only $\xi^+ = 0$). The
parametrization of the $x$ and $p_\sT$ dependent correlator 
becomes\cite{RS-79,MT-96,BM-98} 
\begin{eqnarray}
\Phi(x,\bm{p}_\sT) & = & 
\frac{1}{2}\,\Biggl\{
f_1(x ,\bm p^2_\sT)\,\slsh n_+ + 
f_{1T}^\perp(x ,\bm p^2_\sT)\, \frac{\epsilon_{\mu \nu \rho \sigma}\gamma^\mu 
n_+^\nu p_\sT^\rho S_{\sT}^\sigma}{M}
\nonumber \\ && \mbox{}  
- g_{1s}(x ,\bm p_\sT)\, \slsh n_+ \gamma_5
- h_{1T}(x ,\bm p^2_\sT)\,i\sigma_{\mu\nu}\gamma_5 S_{\sT}^\mu n_+^\nu
\nonumber \\ && \mbox{}
- h_{1s}^\perp(x ,\bm p_\sT)\,\frac{i\sigma_{\mu\nu}\gamma_5 p_\sT^\mu
n_+^\nu}{M} + h_1^\perp (x,\bm p^2_\sT) \, \frac{\sigma_{\mu\nu} p_\sT^\mu
n_+^\nu}{M}\Biggl\}.
\label{paramPhixkt}
\end{eqnarray}
We used the shorthand notation
$g_{1s}(x, \bm p_\sT) \equiv
S_\sL\,g_{1L}(x ,\bm p_\sT^2)
+ \frac{(\bpt\cdot\bm{S}_{\sT})}{M}\,g_{1T}(x ,\bm p_\sT^2)$,
and similarly for $h_{1s}^\perp$.
The parameterization contains two T-odd functions, the
Sivers function $f_{1T}^\perp$ \cite{s90} and the 
function $h_1^\perp$, the distribution function analogue of the 
Collins fragmentation function $H_1^\perp$. The whole
treatment of the fragmentation functions is analogous with dependence on the
quark momentum fraction $z = P_h^-/k^-$ and $k_\sT$. We use capital letters
for the fragmentation functions. At measured $q_\sT$ one deals with a
convolution of two transverse momentum dependent functions, where the
transverse momenta of the partons from different hadrons combine to
$q_\sT$~\cite{RS-79,MT-96,Boer-00}.  A decoupling is achieved by studying cross
sections weighted with the momentum $q_\sT^\alpha$, leaving only the
directional (azimuthal) dependence. The functions that appear in that case are
contained in 
$\Phi_\partial^\alpha (x) \equiv
\int d^2 p_\sT\,\frac{p_\sT^\alpha}{M} \,\Phi(x,\bm p_\sT)$
which projects out the functions in $\Phi(x,\bm p_\sT)$ where $p_\sT$
appears linearly,
\beaq
\Phi_\partial^\alpha (x) & = &
\frac{1}{2}\,\Biggl\{
-g_{1T}^{(1)}(x)\,S_\sT^\alpha\,\slsh n_+\gamma_5
-S_\sL\,h_{1L}^{\perp (1)}(x)
\,\frac{[\gamma^\alpha,\slsh n_+]\gamma_5}{2}
\nonumber \\
&&\quad \mbox{}
-{f_{1T}^{\perp (1)}}(x)
\,\epsilon^{\alpha}_{\ \ \mu\nu\rho}\gamma^\mu n_-^\nu {S_\sT^\rho}
- i\,{h_1^{\perp (1)}}(x)
\,\frac{[\gamma^\alpha, \slsh n_+]}{2}\Biggr\},
\label{Phid}
\eeaq
and transverse moments are defined as
$f^{(n)}(x)$ = $\int d^2p_\sT\,\left(\frac{\bm p_\sT^2}{2M^2}\right)^n
\,f(x,\bm p_\sT)$.

At this point one can invoke Lorentz invariance as a possibility to
rewrite some functions. All functions in $\Phi(x)$ and 
$\Phi_\partial^\alpha(x)$ involve nonlocal matrix elements of two quark 
fields. Before constraining
the matrix elements to the light-cone or lightfront only a limited number
of amplitudes can be written down. This leads to the following
Lorentz-invariance relations
\beaq
&&g_T  = g_1 + \frac{d}{dx}\,g_{1T}^{(1)}, \qquad
h_L = h_1 - \frac{d}{dx}\,h_{1L}^{\perp (1)},
\label{gTrel}
\\
&&f_T =  - \frac{d}{dx}\,f_{1T}^{\perp (1)}, \qquad
\ h =  - \frac{d}{dx}\,h_{1}^{\perp (1)}.
\label{rel4}
\eeaq
From these relations, it is clear that  
the $\bm p_\sT^2/2M^2$ moments of the $p_\sT$-dependent 
functions, appearing in $\Phi_\partial^\alpha(x)$, involve both twist-2
and twist-3 operators. 

Another useful set of functions is obtained as the difference between
the correlator $\Phi_D(x)$ which via equations of motion is connected to
$\Phi^{\rm twist-3}$ and $\Phi_\partial$. This difference corresponds in
$A^+ = 0$ gauge to correlators $\Phi_A$, involving $\langle P,S \vert
\overline \psi_j(0)\,{\cal U}(0,\eta) A_\sT^\alpha (\eta)\,{\cal
U}(\eta,\xi)\,\psi_i(\xi) \vert P,S \rangle$. The difference defines 
interaction-dependent (tilde) functions,
\beaq
&&
x\,g_T(x) - \frac{m}{M}\,h_1(x)-g_{1T}^{(1)}(x) 
+ i\left[ x\,f_T(x) + f_{1T}^{\perp(1)}(x)\right]
\equiv x\,\tilde g_T(x)
+ ix\,\tilde f_T(x),
\label{collinear1}
\\ &&
x\,h_L(x) - \frac{m}{M}\,g_1(x) 
+2\,h_{1L}^{\perp (1)}(x) - ix\,e_L(x)\equiv
x\,\tilde h_L(x) - ix\,\tilde e_L(x),
\\ &&
x\,e(x) - \frac{m}{M}\,f_1(x) + i\left[
x\,h(x) +2\,h_1^{\perp (1)}(x)\right]\equiv
x\,\tilde e(x) + ix\,\tilde h(x).
\label{collinear4}
\eeaq

Using the equations of
motion relations in Eqs.~(\ref{collinear1}) - (\ref{collinear4}) and the
relations based on Lorentz invariance in Eqs.~(\ref{gTrel}) - (\ref{rel4}),
it is straightforward to relate the various twist-3 functions and the
$\bm p_\sT^2/2M^2$ (transverse) moments of $p_\sT$-dependent
functions. The results e.g. for the $h$-functions are (omitting quark mass
terms) are
\beaq
&&
h_L(x) = 
2x\int_x^1 dy\ \frac{h_1(y)}{y^2}
+ \left[ \tilde h_L(x) - 2x\int_x^1 dy\ \frac{\tilde h_L(y)}{y^2}\right],
\\ &&
\frac{h_{1L}^{\perp(1)}(x)}{x^2} =
-\int_x^1 dy\ \frac{h_1(y)}{y^2}
+ \int_x^1 dy\ \frac{\tilde h_L(y)}{y^2},
\\ &&
h(x) = 
\left[ \tilde h(x) - 2x\int_x^1 dy\ \frac{\tilde h(y)}{y^2}\right],
\\ &&
\frac{h_{1}^{\perp(1)}(x)}{x^2} =
\int_x^1 dy\ \frac{\tilde h(y)}{y^2} .
\eeaq
Actually, we need not consider the T-odd functions separately. They can be
simply considered as imaginary parts of other functions, when we allow complex
functions. In particular one can expand the correlation functions into
matrices in Dirac space~\cite{BBHM} to show that the relevant combinations are 
$(g_{1T} - i\,f_{1T}^{\perp})$ which we can treat together as one complex 
function $g_{1T}$. Similarly we can use complex functions
$(h_{1L}^\perp + i\,h_1^\perp)$ $\rightarrow$ $h_{1L}^\perp$,
$(g_T + i\,f_T)$ $\rightarrow$ $g_T$,
$(h_L + i\,h)$ $\rightarrow$ $h_L$,
$(e + i\,e_L)$ $\rightarrow$ $e$. The functions $f_1$, $g_1$ and $h_1$ remain
real, they don't have T-odd partners. 

As mentioned the evolution of the twist-2 functions and the tilde functions in
known. The twist-2 functions have an autonomous evolution of the form
\beq
\frac{d}{d\tau} \,f(x,\tau) = \frac{\alpha_s(\tau)}{2\pi}
\,\int_x^1 \frac{dy}{y} \ P^{[f]}\left(\frac{x}{y}\right)\,f(y,\tau),
\eeq
where $\tau$ = $\ln Q^2$ and $P^{[f]}$ are the splitting functions. 
In the large $N_c$ limit, also the tilde functions have an autonomous
evolution. Using the relations given above, we then find the evolution of the
transverse moments,
\beaq
&&
\frac{d}{d\tau}\,g_{1T}^{(1)}(x,\tau) 
= \frac{\alpha_s(\tau)}{4\pi}\,N_c\int_x^1 dy\,\Biggl\{
\left[\frac{1}{2}\,\delta(y-x) + \frac{x^2+xy}{y^2(y-x)_+}\right]
\,g_{1T}^{(1)}(y,\tau)
\nonumber \\ && \hspace{8cm}
+ \frac{x^2}{y^2}\,g_1(y,\tau)\Biggr\} ,
\\ && 
\frac{d}{d\tau}\,h_{1L}^{\perp (1)}(x,\tau) 
= \frac{\alpha_s(\tau)}{4\pi}\,N_c\int_x^1 dy\,\Biggl\{
\left[\frac{1}{2}\,\delta(y-x) + \frac{3x^2-xy}{y^2(y-x)_+}\right]
\,h_{1L}^{\perp (1)}(y,\tau)
\nonumber \\ && \hspace{8cm}
-\frac{x}{y}\,h_1(y,\tau)\Biggr\}.
\eeaq
Next we note that apart from a
$\gamma_5$ matrix the operator structures of the T-odd functions 
$f_{1T}^{\perp (1)}$ and $h_1^{\perp (1)}$ are in fact the same 
as those of $g_{1T}^{(1)}$ and $h_{1L}^{\perp (1)}$ (or as mentioned before,
they can be considered as the imaginary part of these functions~\cite{BBHM}).
This implies that for the non-singlet functions, one immediately obtains the
(autonomous) evolution of these T-odd functions. In particular we obtain for
the Collins fragmentation function (at large $N_c$),
\beq
\frac{d}{d\tau} \,z H_{1}^{\perp (1)}(z,\tau) 
= \frac{\alpha_s}{4\pi}\; N_c\; \int_z^1 dy\, \left[
\frac{1}{2}\,\delta(y-z) + \frac{3y-z}{y(y-z)_+} \right] 
\,y H_{1}^{\perp (1)}(y,\tau),
\eeq
which should prove useful for the comparison of data on Collins
function asymmetries from  different experiments, performed at different
energies.

Summarizing, we have obtained evolution equations of the $p_\sT$-dependent
functions that appear in asymmetries and that are not suppressed by explicit
powers of the hard momentum. But as functions of transverse momentum they are
not of  definite twist


\end{document}